\documentclass{ws-p8-50x6-00}

\def\lsim{\mathrel{\rlap{\lower 4pt \hbox{\hskip 1pt $\sim$}}\raise 1pt \hbox
        {$<$}}}
\def\gsim{\mathrel{\rlap{\lower 4pt \hbox{\hskip 1pt $\sim$}}\raise 1pt \hbox
        {$>$}}}

\begin{document}

\title{Hypernova Nucleosynthesis and Implications
for Cosmic Chemical Evolution}

\author{Takayoshi Nakamura$^{1}$, Hideyuki Umeda$^{1}$,
Koichi Iwamoto$^{2}$, Ken'ichi Nomoto$^{1}$,
Masa-aki Hashimoto$^{3}$, W. Raphael Hix$^{4}$, 
and Friedrich-Karl Thielemann$^{5}$}

\address{
$^{1}$Department of Astronomy, School of Science,
University of Tokyo, Tokyo, Japan\\
$^{2}$Dept. of Physics, College of Science and Technology,
Nihon Univ., Tokyo, Japan\\
$^{3}$Department of Physics, School of Sciences,
University of Kyushu, Fukuoka, Japan\\
$^{4}$Department of Physics and Astronomy, University of 
Tennessee, Knoxville, TN
and Physics Division, Oak Ridge National Laboratory, Oak Ridge, TN, USA\\
$^{5}$Department f\"ur Physik und Astronomie,
Universit\"at Basel, Switzerland\\
}


\maketitle

\abstracts{
We examine the characteristics of nucleosynthesis in 'hypernovae',
i.e., supernovae with very large explosion energies
($ \gsim 10^{52} $ ergs).
Implications for the cosmic chemical evolution and the abundances
in M82 are discussed.
}

\section{
Enhancement of [Fe/O], [Ti/O], and [Si, S/O]
}
The explosion energy of SN1998bw is estimated to be $E$ = 3 - 6$
\times 10^{52}$ ergs, which is about thirty
times larger than that of a normal supernova (Iwamoto et al. 1998;
Woosley et al. 1999; Nakamura et al. 2001).
We explore nucleosynthesis in such energetic
core-collapse supernovae, 
called 'hypernovae' (Nakamura et al. 2000),
and find the following characteristics.
In hypernovae, both complete and incomplete Si-burning take place in
more extended, and hence, lower density regions, so that the
$\alpha$-rich freezeout is enhanced in comparison with normal supernova
nucleosynthesis.  Thus $^{44}$Ca ($\leftarrow$ $^{44}$Ti)
and $^{48}$Ti are produced more
abundantly than in canonical supernovae.  Oxygen and carbon burning
also takes place in more extended regions for the larger explosion
energy. Therefore, the fuel elements O, C, Al are less abundant
while a larger amount of burning products such as Si, S, and Ar are
synthesized by oxygen burning.
In short, [Ti/O] and [Si, S, Ar/O]
as well as [Fe/O] are enhanced in hypernovae.

\section{Comparison with abundances in metal-poor stars}

Ti is known to be deficient in Galactic chemical
evolution models that use supernova yields currently available (e.g.,
Timmes et al. 1995; Thielemann et al. 1996),
especially at [Fe/H] $\lsim -1$, when Type Ia supernovae (SNe Ia)
have not contributed to the Galactic chemical evolution.
However, if the contribution from hypernovae
is relatively large,
this problem could be relaxed, because, as we have seen,
the $\alpha$-rich freezeout is enhanced in hypernovae
and Ti is produced more abundantly than normal supernovae.

Another feature of hypernova nucleosynthesis is a large amount of Fe.
One hypernova can produce
2 - 10 times more Fe than normal core-collapse supernovae.
This large iron production leads to small ratios of $\alpha$-elements over
iron in hypernovae.  In this connection,
the abundance pattern of the very metal-poor binary CS22873-139
([Fe/H] = $-3.4$) is interesting.  This binary has only an upper limit
to [Sr/Fe] $< -1.5$, and therefore was suggested to be a second
generation star (Nordstr\"om et al. 2000; Spite et al. 2000).  The
interesting pattern is that this binary shows almost solar Mg/Fe and
Ca/Fe ratios, as is the case with hypernovae.
Another feature of CS22873-139 is enhanced Ti/Fe ([Ti/Fe]
$\sim + 0.6$; Nordstr\"om et al. 2000; Spite et al. 2000),
which could be explained by a hypernova explosion.

\section{Abundances in the starburst galaxy M82}
X-ray emissions from the starburst galaxy M82 were observed with ASCA
and the abundances of several heavy elements were measured (Tsuru et
al. 1997).  Tsuru et al. (1997) found that the overall metallicity of
M82 is quite low, i.e., O/H and Fe/H are only 0.06 - 0.05 times solar,
while Si/H and S/H are $\sim$ 0.40 - 0.47 times solar.  This implies
that the abundance ratios are peculiar, i.e., the ratio O/Fe is about
solar, while the ratios of Si and S relative to O and Fe are as high
as $\sim$ 6 - 8.  These ratios are very different from those ratios
in Type II supernovae (SNe II).
The age of M82 is estimated to be $\lsim 10^8$ years, which
is too young for SNe Ia to contribute to enhance Fe
relative to O.  Tsuru et al. (1997) also estimated that the explosion
energy required to produce the observed amount of hot plasma per
oxygen mass is significantly larger than that of normal SNe II (here
the oxygen mass dominates the total mass of the heavy elements).
Tsuru et al. (1997) thus concluded that neither SN Ia nor SN II can
reproduce the observed abundance pattern of M82.

Compared with normal SNe II, the important characteristic of hypernova
nucleosynthesis is the large Si/O, S/O, and Fe/O ratios.
Figure 1 shows the good agreement between the hypernova model
and the observed abundances in M82 (Umeda et al. 2001).
Hypernovae could also produce larger $E$ per oxygen
mass than normal SNe II.  We therefore suggest that hypernova
explosions may make important contributions to the metal enrichment
and energy input to the interstellar matter in M82.  If the IMF of the
star burst is relatively flat compared with Salpeter IMF, the
contribution of very massive stars and thus hypernovae could be much
larger than in our Galaxy.

\begin{figure}[t]
\hspace{1.2cm}
\epsfxsize=18.5pc 
\epsfbox{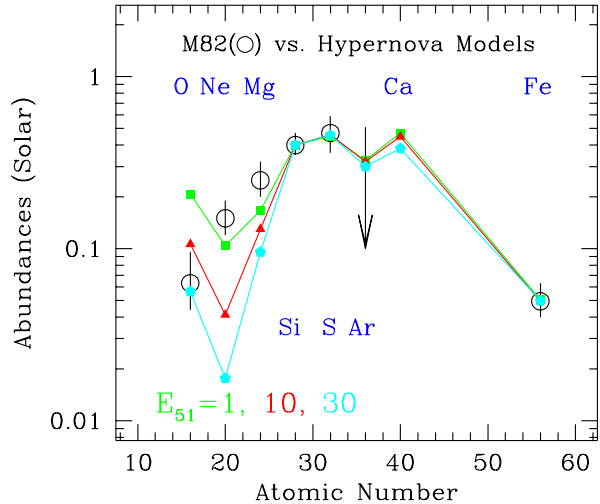} 
\caption{
Abundance patterns in the ejecta of 25$M_\odot$ metal-free
SN II and hypernova models
compared with abundances (relative to the solar values)
of M82 observed with ASCA (Tsuru
et al. 1997).  Here, the open circles with error bars show the M82 data.
The filled square, triangle, and pentagons represent $E_{51}$=1, 10, and
30 models, respectively, where $E_{51}$ is the explosion energy in
$10^{51}$ ergs.  Theoretical abundances are normalized to the observed
Si data, and the mass cuts are chosen to eject 0.07, 0.095, and 0.12
($M_\odot$) Fe for $E_{51}$ = 1, 10, and 30,
respectively (Umeda et al. 2001).
}
\end{figure}

\end{document}